\documentclass[12pt]{article}

\usepackage{graphicx}
\usepackage{epsfig}

\newcommand{\Z}{{\sf Z \!\!\! Z}}

\makeatletter
\@addtoreset{equation}{section}
\makeatother

\begin{document}
 
\title{Efficient Cluster Algorithm for $CP(N-1)$ Models} 
\author{B.\ B.\ Beard$^a$, M.\ Pepe$^{b,c}$, S.\ Riederer$^c$, 
and U.-J.\ Wiese$^c$
\\ \\
$^a$ Department of Physics and Mechanical Engineering \\ 
Christian Brothers University, Memphis, TN 38104, U.S.A. \\ \\
$^b$ Istituto Nazionale di Fisica Nucleare and \\
Dipartimento di Fisica, Universit\`a di Milano-Bicocca \\
3 Piazza della Scienza, 20126 Milano, Italy \\ \\
$^c$ Institute for Theoretical Physics \\ 
Bern University, Sidlerstrasse 5, 3012 Bern, Switzerland}

\maketitle

\begin{abstract}

Despite several attempts, no efficient cluster algorithm has been constructed 
for $CP(N-1)$ models in the standard Wilson formulation of lattice field 
theory. In fact, there is a no-go theorem that prevents the construction of an 
efficient Wolff-type embedding algorithm. In this paper, we construct an 
efficient cluster algorithm for ferromagnetic $SU(N)$-symmetric quantum spin 
systems. Such systems provide a regularization for $CP(N-1)$ models in the 
framework of D-theory. We present detailed studies of the autocorrelations and 
find a dynamical critical exponent that is consistent with $z = 0$.

\end{abstract}

\newpage

\section{Introduction}

Cluster algorithms are among the most efficient tools for numerical simulations
of nonperturbative lattice field theories and statistical mechanics models. The
first cluster algorithm for discrete classical spins was developed by Swendsen 
and Wang for Ising and Potts models \cite{Swe87}. This algorithm was 
generalized by Wolff to $O(N)$ models formulated in terms of continuous 
classical spins \cite{Wol89}. In the 2-d $O(3)$ model the Wolff cluster 
algorithm has been extended to a meron-cluster algorithm which has been used to
solve the complex action problem at non-zero $\theta$-vacuum angle 
\cite{Bie95}. $CP(N-1)$ models in two dimensions \cite{DAd78,Eic78,Gol78} 
provide a generalization of the $O(3) = CP(1)$ model alternative to $O(N)$ 
models. They have a global 
$SU(N)$ symmetry and share several features with QCD, including asymptotic 
freedom, the nonperturbative dynamical generation of a mass-gap, a topological 
charge, and thus nontrivial $\theta$-vacua. Finding efficient algorithms for 
$CP(N-1)$ models is difficult in the standard Wilson formulation. In 
particular, a Wolff-type embedding cluster algorithm turned out to be 
inefficient for $CP(N-1)$ models \cite{Jan92}. In fact, there is even a no-go 
theorem that prevents the success of this type of algorithms \cite{Car93}. 
Still, a rather efficient multigrid algorithm does exist \cite{Has92,Has92a} 
for $\theta = 0$. However, due to a very severe complex action problem, no 
efficient algorithm has ever been found for non-zero vacuum angle. 
Consequently,  in the Wilson formulation numerical studies of $\theta$-vacua 
\cite{Wie89,Bur01} have been limited to moderate volumes, or are based on 
additional assumptions \cite{Azc02,Azc03,Azc04,Ima04}.

{\em D-theory} is an alternative formulation of field theory in which 
continuous classical fields arise dynamically as collective excitations by the 
{\em dimensional} reduction of {\em discrete} variables such as quantum spins 
\cite{Cha97,Bro99,Bro04}. The discrete nature of the quantum variables allows 
the development of new types of cluster algorithms. The first cluster 
algorithm for quantum spins was developed in \cite{Wie92}. This algorithm works
efficiently for 1-d ferro- and antiferromagnetic quantum spin chains, while the
loop-cluster algorithm \cite{Eve93,Eve03} works efficiently in any dimension. 
For example, the loop-algorithm has led to very accurate simulations of the 
2-d Heisenberg quantum antiferromagnet \cite{Wie94}, which can even be 
performed in continuous Euclidean time \cite{Bea96}. The D-theory formulation 
of field theory has been used, for example, to construct an efficient cluster 
algorithm that solves the complex action problem in the 2-d $O(3)$ model at 
non-zero chemical potential \cite{Cha02}. In this paper we construct an
efficient algorithm for the D-theory formulation of $CP(N-1)$ models. The 
D-theory approach allows efficient simulations both at $\theta = 0$ and $\pi$.
In D-theory $CP(N-1)$ models are represented by ferro- or antiferromagnetic 
$SU(N)$ quantum spin systems. Just like the $SU(2)$ Heisenberg antiferromagnet,
$SU(N)$ quantum antiferromagnets can be simulated with the loop-cluster 
algorithm \cite{Har03} which was recently reviewed in \cite{Kaw03,Tro03}. Here 
we construct an $SU(N)$ loop-cluster algorithm for a ferromagnet to study 
$CP(N-1)$ models at $\theta = 0$, and we investigate their numerical 
efficiency. We find a dynamical exponent for critical slowing down which is
consistent with $z = 0$. The D-theory method even solves the $\theta$-vacuum
sign problem, and has recently allowed us to confirm the conjectured first 
order phase transition \cite{Sei84,Aff88} at $\theta = \pi$ \cite{Bea04}.

The paper is organized as follows. We present the standard Wilson formulation
of lattice $CP(N-1)$ models in section 2 and the D-theory formulation in 
section 3. Section 4 describes the path integral formulation for $CP(N-1)$
models in terms of ferromagnetic quantum spins transforming in the fundamental
representation of $SU(N)$. The corresponding loop-cluster algorithm is
constructed in section 5, and its efficiency is investigated in section 6.
Finally, section 7 contains our conclusions.

\section{Standard Formulation of $CP(N-1)$ Models}

The manifold $CP(N-1) = SU(N)/U(N-1)$ is a $(2N-2)$-dimensional coset space 
relevant in the context of the spontaneous breakdown of an $SU(N)$ symmetry to 
a $U(N-1)$ subgroup. In particular, in more than two space-time dimensions 
($d > 2$) the corresponding Goldstone bosons are described by an $N \times N$ 
matrix-valued field $P(x) \in CP(N-1)$ which obeys
\begin{equation}
P(x)^2 = P(x), \quad P(x)^\dagger = P(x), \quad \mbox{Tr} P(x) = 1.
\end{equation}
For $d = 2$ the Hohenberg-Mermin-Wagner-Coleman theorem implies that the 
$SU(N)$ symmetry cannot break spontaneously. Correspondingly, similar to 4-d
non-Abelian gauge theories, the field $P(x)$ develops a mass-gap 
nonperturbatively. Motivated by these observations, D'Adda, Di Vecchia, and
L\"uscher~\cite{DAd78} introduced $CP(N-1)$ models as interesting toy models
for QCD. The corresponding Euclidean action is given by
\begin{equation}
\label{CPNaction}
S[P] = \int d^2x \ \frac{1}{g^2} \mbox{Tr}[\partial_\mu P \partial_\mu P] -
i \theta Q[P],
\end{equation}
where $g^2$ is the dimensionless coupling constant. The topological charge
\begin{equation}
\label{CPNcharge}
Q[P] = \frac{1}{2 \pi i} \int d^2x \ \epsilon_{\mu\nu} \mbox{Tr}
[P \partial_\mu P \partial_\nu P] \in \Pi_2[CP(N-1)] = \Z,
\end{equation}
takes integer values in the second homotopy group of $CP(N-1)$\footnote{Note 
that the factor $i$ in eq.(\ref{CPNcharge}) is necessary for obtaining a 
real-valued quantity.}, and $\theta$ is the corresponding vacuum angle. Note 
that both the action and the topological charge are invariant under global 
transformations $\Omega \in SU(N)$ 
\begin{equation}
P(x)' = \Omega P(x) \Omega^\dagger.
\end{equation}

The topological term explicitly breaks the charge
conjugation symmetry $C$ which acts as $^CP(x) = P(x)^*$, such that
$Q[^CP] = - Q[P]$. It should be noted that $C$ is not explicitly broken at
$\theta = 0$ or $\pi$. In particular, for $\theta = \pi$ the Boltzmann weight
$\exp(i \theta Q[P]) = (-1)^{Q[P]}$ is $C$-invariant. For $N \geq 3$ it has 
been conjectured that there is a first order phase transition at 
$\theta = \pi$, at which charge conjugation is broken 
spontaneously~\cite{Sei84}. Recently, this has been confirmed numerically using
a cluster algorithm similar to the one described below~\cite{Bea04}. For 
$CP(1) = O(3)$, on the other hand, charge conjugation remains unbroken, but 
there is still a second order phase transition at 
$\theta = \pi$~\cite{Hal83,Aff86,Bie95}. At that point the model corresponds to
the $k = 1$ Wess-Zumino-Novikov-Witten conformal field 
theory~\cite{Wes71,Nov81,Wit84}.

It is straightforward to define $CP(N-1)$ models beyond perturbation theory in 
the framework of Wilson's lattice regularization. Then the field $P_x \in 
CP(N-1)$ is defined on the sites $x$ of a quadratic lattice and the standard 
lattice action is given by
\begin{equation}
S[P] = - \frac{2}{g^2} \sum_{x,\mu} \mbox{Tr}[P_x P_{x+\hat \mu}],
\end{equation}
where $\hat \mu$ is the unit-vector in the $\mu$-direction.

\section{D-Theory Formulation of $CP(N-1)$ Models}

In this section we describe an alternative formulation of field theory in which
the $d$-dimensional $CP(N-1)$ model emerges from the dimensional reduction of 
discrete variables --- in this case $SU(N)$ quantum spins in $d+1$ space-time 
dimensions. The dimensional reduction of discrete variables is the 
key ingredient of D-theory, which provides an alternative nonperturbative
lattice regularization of field theory. In D-theory we start from a 
ferromagnetic system of $SU(N)$ quantum spins located at the sites $x$ of a 
$d$-dimensional periodic hypercubic lattice of size $L^d$. The $SU(N)$ spins 
are represented by Hermitean operators $T_x^a = \frac{1}{2} \lambda_x^a$ that 
generate the group $SU(N)$ (e.g.\ the Gell-Mann matrices for the triplet 
representation of $SU(3)$) and thus they obey
\begin{equation}
[T_x^a,T_y^b] = i \delta_{xy} f_{abc} T_x^c, \quad
\mbox{Tr}(T_x^a T_y^b) = \frac{1}{2} \delta_{xy} \delta_{ab}.
\end{equation}
In principle, these generators can be taken in any irreducible representation 
of $SU(N)$. However, not all representations lead to 
spontaneous symmetry breaking from $SU(N)$ to $U(N-1)$ and thus to $CP(N-1)$ 
models. The Hamilton operator for an $SU(N)$ ferromagnet takes the form
\begin{equation}
H = - J \sum_{x,i} T_x^a T_{x+\hat i}^a,
\end{equation}
where $J>0$ is the exchange coupling. By construction, the Hamilton operator is
invariant under the global $SU(N)$ symmetry, i.e.\ it commutes with the total 
spin given by 
\begin{equation}
T^a = \sum_x T_x^a.
\end{equation}
The Hamiltonian $H$ describes the evolution of the quantum spin system in an 
extra dimension of finite extent $\beta$. In D-theory this extra dimension is 
not the Euclidean time of the target theory, which is part of the 
$d$-dimensional lattice. Instead, it is an additional compactified dimension 
which ultimately disappears via dimensional reduction. The quantum partition 
function
\begin{equation}
Z = \mbox{Tr} \exp(- \beta H)
\end{equation}
(with the trace extending over the Hilbert space) gives rise to periodic 
boundary conditions in the extra dimension.

For $d \geq 2$ the ground state of the quantum spin system has a broken global 
$SU(N)$ symmetry. The choice of the $SU(N)$ representation determines the
symmetry breaking pattern. We choose a totally symmetric $SU(N)$ representation
corresponding to a Young tableau with a single row containing $n$ boxes. It is 
easy to construct the ground states of the $SU(N)$ ferromagnet, and one finds 
spontaneous symmetry breaking from $SU(N)$ to $U(N-1)$. Consequently, there are
$(N^2 - 1) - (N-1)^2 = 2N - 2$ massless Goldstone bosons described by a field 
$P(x)$ in the coset space $SU(N)/U(N-1) = CP(N-1)$. In the leading order of 
chiral perturbation theory the Euclidean action for the Goldstone boson field 
is given by
\begin{equation}
\label{ferroaction}
S[P] = \int_0^\beta dt \int d^dx \ \mbox{Tr}
[\rho_s \partial_\mu P \partial_\mu P - \frac{2 n}{a^2} \int_0^1 d\tau \ 
P \partial_t P \partial_\tau P].
\end{equation}
Here $\rho_s = J n^2/4$ is the spin stiffness which is analogous to the pion 
decay constant in QCD. The second term in eq.(\ref{ferroaction}) is a 
Wess-Zumino
term which involves an integral over an interpolation parameter $\tau$. The 
point $\tau = 1$ corresponds to the physical space-time, such that 
$P(x,t,1) = P(x,t)$. At $\tau = 0$, on the other hand, the field takes the 
constant value $P(x,t,0) = \mbox{diag}(1,0,...,0)$. At intermediate values of 
$\tau$ the field is smoothly interpolated between the two limiting cases. There
is no topological obstruction against such an interpolation because 
$\Pi_1[CP(N-1)]$ is trivial. The integrand in the Wess-Zumino term is a 
total derivative. Hence, the integral depends only on the boundary values  at 
$\tau = 1$, i.e.\ on the values of the field in the physical space-time. 
However, since $\Pi_2[CP(N-1)] = \Z$, there is an integer ambiguity of the 
integral depending on which particular interpolation is chosen for 
$P(x,t,\tau)$. In order to make sure that this ambiguity does not affect the 
physics, the prefactor of the Wess-Zumino term must be quantized. 
Remarkably, this prefactor is just determined by the number of boxes $n$ of the
chosen $SU(N)$ representation. 

Let us first consider the $d = 2$ case. For $\beta = \infty$ the system then
has a spontaneously broken global symmetry and thus massless Goldstone bosons.
However, as soon as $\beta$ becomes finite, due to the 
Hohenberg-Mermin-Wagner-Coleman theorem, the symmetry can no longer be broken,
and, consequently, the Goldstone bosons pick up a small mass $m$
nonperturbatively. As a result, the corresponding correlation length 
$\xi = 1/m$ becomes finite and the $SU(N)$ symmetry is restored over that 
length scale. The question arises if $\xi$ is bigger or smaller than the
extent $\beta$ of the extra dimension. When $\xi \gg \beta$ the Goldstone boson
field is essentially constant along the extra dimension and the system 
undergoes dimensional reduction. Since the Wess-Zumino term contains 
$\partial_t P$, it simply vanishes after dimensional reduction. 
Correspondingly, the action dimensionally reduces to
\begin{equation}
\label{targetaction}
S[P] = \beta \rho_s \int d^2x \ \mbox{Tr}[\partial_\mu P \partial_\mu P],
\end{equation}
which is just the action of the 2-d target $CP(N-1)$ model at $\theta=0$. The
coupling constant of the 2-d model is determined by the extent of the extra
dimension and is given by
\begin{equation}
\frac{1}{g^2} = \beta \rho_s.
\end{equation}
Due to asymptotic freedom of the 2-d $CP(N-1)$ model, for small $g^2$ the 
correlation length is exponentially large, i.e.\
\begin{equation}
\xi \propto \exp(4 \pi \beta \rho_s/N).
\end{equation}
Here $N/4 \pi$ is the 1-loop coefficient of the perturbative $\beta$-function.
Indeed, one sees that $\xi \gg \beta$ as long as $\beta$ itself is sufficiently
large. In particular, somewhat counter-intuitively, dimensional reduction 
happens in the large $\beta$ limit because $\xi$ then grows exponentially. In
D-theory one approaches the continuum limit not by varying a bare coupling 
constant but by increasing the extent $\beta$ of the extra dimension. This
mechanism of dimensional reduction of discrete variables is generic and occurs
in all asymptotically free D-theory models \cite{Cha97,Bro99,Bro04}. It should
be noted that (just like in the standard approach) no fine-tuning is needed
to approach the continuum limit.

In order to show that D-theory indeed defines the same field theory in the
continuum limit as the standard Wilson approach, we have evaluated a universal
physical quantity in both regularizations \cite{Bea04}. A convenient quantity 
is the universal finite-size scaling function $F(y) = \xi(2 L)/\xi(L)$. Here 
$\xi(L)$ is the correlation length (obtained 
with the second moment method) in a finite system of size $L$, and $y =
\xi(L)/L$ is a finite size scaling variable that measures the size of the
system in physical units.
\begin{figure}[tb]
\begin{center}
\epsfig{file=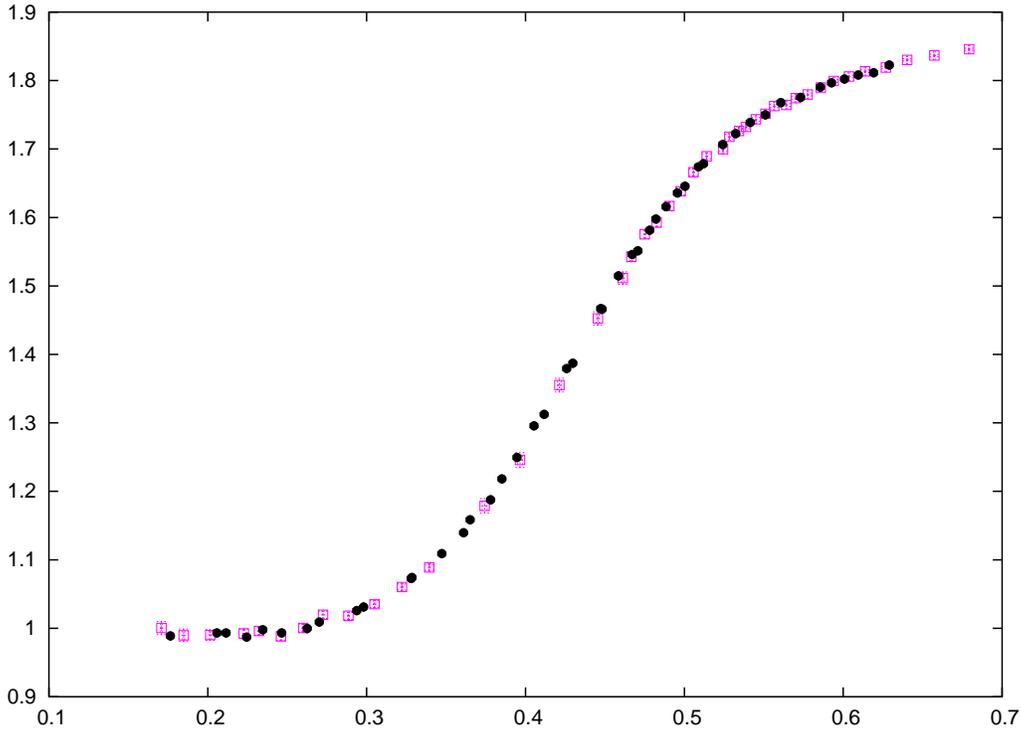,width=10cm,angle=-90}
\vspace{0.2cm}
\caption{\it Monte Carlo data for the universal finite-size scaling function 
$F(y)$ of the 2-d $CP(2)$ model. The filled circles represent D-theory data
from a $(2+1)$-d $SU(3)$ quantum ferromagnet at $\beta J = 6$, while the open
squares correspond to data obtained with the standard Wilson 2-d lattice field
theory at $1/g^2 = 2.25$.}
\end{center}
\end{figure}
Figure 1 shows Monte Carlo data for $F(y)$ obtained both from D-theory and from
the standard Wilson approach. Up to small scaling violations, the agreement of 
the two data sets confirms that, after dimensional reduction, the $(2+1)$-d 
$SU(N)$ quantum ferromagnet indeed provides the correct continuum limit of the 
2-d $CP(N-1)$ model. Thanks to the cluster algorithm, the D-theory framework 
allows calculations that are much more accurate than the ones using Wilson's 
approach.

Let us also discuss the $d > 2$ case. Then dimensional reduction still occurs,
but the Goldstone bosons remain massless even at finite $\beta$. However, if 
$\beta$ is decreased sufficiently, the $SU(N)$ symmetry is restored and the
Goldstone modes are replaced by massive excitations. By fine-tuning $\beta$ to
the critical value (again, as in the standard approach) one can reach a 
continuum limit of the target theory, provided that the corresponding phase 
transition is second order.

\section{Path Integral Representation of $SU(N)$ Quantum Spin Systems}

Let us construct a path integral representation for the partition function $Z$
of the $SU(N)$ quantum spin ferromagnet introduced above. In an intermediate 
step we introduce a lattice in the 
Euclidean time direction, using a Trotter decomposition of the Hamiltonian.
However, since we are dealing with discrete variables, the path integral is
completely well-defined even in continuous Euclidean time. Also the cluster 
algorithm to be described in the following section can operate directly in the
Euclidean time continuum \cite{Bea96}. Hence, the final results are completely 
independent of the Trotter decomposition. In $d$ spatial dimensions (with an 
even extent $L$) we decompose the Hamilton operator into $2d$ terms
\begin{equation}
H = H_1 + H_2 + ... + H_{2d},
\end{equation}
with
\begin{equation}
H_i = \!\! \sum_{\stackrel{x = (x_1,x_2,...,x_d)}{x_i \rm{even}}} \!\! 
h_{x,i}, \ \
H_{i+d} = \!\! \sum_{\stackrel{x = (x_1,x_2,...,x_d)}{x_i \rm{odd}}} \!\! 
h_{x,i}.
\end{equation}
The individual contributions
\begin{equation}
h_{x,i} = - J \ T_x^a T_{x+\hat i}^a,
\end{equation}
to a given $H_i$ commute with each other, but two different $H_i$ do not 
commute. Using the Trotter formula, the partition function then takes the form
\begin{eqnarray}
Z\!\!\!&=&\!\!\!\lim_{M \rightarrow \infty} \! \mbox{Tr} 
\left\{\exp(- \epsilon H_1) \exp(- \epsilon H_2) ...
\exp(- \epsilon H_{2d})\right\}^M.
\end{eqnarray}
We have introduced $M$ Euclidean time-slices with $\epsilon = \beta/M$ being 
the lattice spacing in the Euclidean time direction. 

Inserting complete sets of spin states $q \in \{u,d,s,...\}$ the partition 
function takes the form
\begin{equation}
Z = \sum_{[q]} \exp(- S[q]).
\end{equation}
The sum extends over configurations $[q]$ of spins  $q(x,t)$ on a 
$(d+1)$-dimensional space-time lattice of points $(x,t)$. The Boltzmann factor
\begin{eqnarray}
\exp(- S[q])\!\!\!\!&=&\!\!\!\!\!\!
\prod_{p=0}^{M-1} \prod_{i=1}^{d} \!\!\!\!
\prod_{\stackrel{x = (x_1,x_2,...,x_d)}
{x_i even,~t = 2dp+i-1}} \!\!\!\!\!\!\!\!\!\!\!\!
\exp\{- s[q(x,t),q(x+\hat i,t),q(x,t+1),q(x+\hat i,t+1)]\} \nonumber \\
&&\times\!\!\!\!\!\!\!\!\!\! \prod_{\stackrel{x = (x_1,x_2,...,x_d)}
{x_i odd,~t = 2dp+d+i-1}} \!\!\!\!\!\!\!\!\!\!\!\!\!
\exp\{- s[q(x,t),q(x+\hat i,t),q(x,t+1),q(x+\hat i,t+1)]\},
\end{eqnarray}
is a product of space-time plaquette contributions with
\begin{eqnarray}
\label{Boltzmannf}
&&\exp(- s[u,u,u,u]) = \exp(- s[d,d,d,d]) = 1,
\nonumber \\
&&\exp(- s[u,d,u,d]) = \exp(- s[d,u,d,u]) = 
\frac{1}{2}[1 + \exp(- \epsilon J)],
\nonumber \\
&&\exp(- s[u,d,d,u]) = \exp(- s[d,u,u,d]) =
\frac{1}{2}[1 - \exp(- \epsilon J)].
\end{eqnarray}
In these expressions the flavors $u$ and $d$ can be permuted to other values. 
All the other Boltzmann factors are zero, which implies several constraints on 
allowed configurations. 

In the following we will consider the uniform magnetization
\begin{equation}
{\cal M} = \sum_{x,t} [\delta_{q(x,t),u} - \delta_{q(x,t),d}],
\end{equation}
which represents one component of $T^a = \sum_x T^a_x$, as well as the
corresponding susceptibility
\begin{equation}
\label{sus}
\chi = \frac{1}{\beta L^d} {\langle {\cal M}^2 \rangle},
\end{equation}
where $L^d$ is the spatial volume of the system.

\section{Cluster Algorithm for $SU(N)$ Quantum Ferromagnets}

Let us now discuss the cluster algorithm for the $SU(N)$ quantum ferromagnet.
Just like the original $SU(2)$ loop-cluster algorithm \cite{Eve93,Wie94}, the
$SU(N)$ cluster algorithm builds a closed loop connecting neighboring lattice 
points with the spin in the same quantum state, and then changes the state of
all spins in that cluster to a different randomly chosen common value. To begin
cluster 
growth, an initial lattice point $(x,t)$ is picked at random. The spin located
at that point participates in two plaquette interactions, one before and one 
after $t$. One picks one interaction arbitrarily and considers the states
of the other spins on that plaquette. One of the corners of this interaction
plaquette will be the next point on the loop. For configurations
$C_1 = [u,d,u,d]$ or $[d,u,d,u]$ the next point is the time-like neighbor of
$(x,t)$ on the plaquette, while for configurations $C_2 = [u,d,d,u]$ or 
$[d,u,u,d]$ the next point is the diagonal neighbor. If the states are all the
same, i.e.\ for $C_3 = [u,u,u,u]$ or $[d,d,d,d]$, with probability 
\begin{equation}
p = \frac{1}{2}[1 + \exp(-\epsilon J)]
\end{equation}
the next point on the loop is again the time-like neighbor, and with 
probability $(1 - p)$ it is the diagonal neighbor. The next point on the loop
belongs to another interaction plaquette on which the same process is 
repeated. In this way the loop grows until it finally closes. The cluster 
rules are consistent with detailed balance, i.e.\
\begin{equation}
p(C_i) w(C_i \rightarrow C_j) = p(C_j) w(C_j \rightarrow C_i),
\end{equation}
where $p(C_i) = \exp(- s[C_i])$ is the Boltzmann weight of the plaquette
configuration and $w(C_i \rightarrow C_j)$ is the transition probability to go
from $C_i$ to $C_j$.

With this single-cluster algorithm, the uniform susceptibility of 
eq.(\ref{sus}) has an improved estimator
\begin{equation}
\chi = \langle |{\cal C}| \rangle,
\end{equation}
given in terms of the cluster size
\begin{equation}
|{\cal C}| = \frac{\beta}{2d M} \sum_{(x,t) \in {\cal C}} 1.
\end{equation}
This shows explicitly that the clusters are physical objects whose size is
directly related to a physical quantity.

\section{Efficiency of the Algorithm in the Continuum Limit}

In order to quantify the efficiency of our numerical method, we have used a
multi-cluster algorithm for a $(2+1)$-dimensional $SU(3)$ ferromagnet in 
discrete Euclidean time to investigate the autocorrelation times of the uniform
magnetization which gives the cleanest signal. A Monte Carlo history of this 
observable is shown in figure 2. This already indicates that autocorrelations 
are very much suppressed. 
\begin{figure}[tb]
\begin{center}
\vspace{-0.4cm}
\epsfig{file=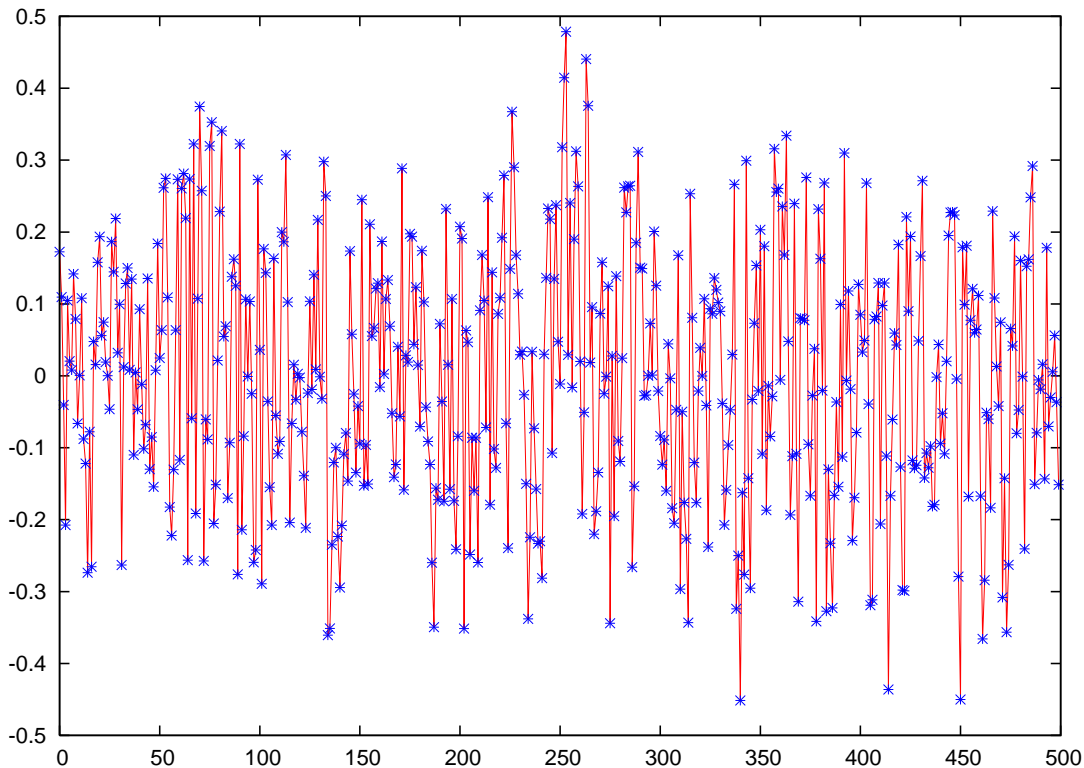,width=15cm}
\end{center}
\caption{\it History of the magnetization of an $SU(3)$ ferromagnet on a
$320^2$ lattice at $\beta J = 6.75$ as a function of the Monte Carlo time 
(measured in units of sweeps with the multi-cluster algorithm).}
\end{figure}
\begin{table}[hbt]
\begin{center}
\begin{tabular}{|c|c|c|c|c|c|c|}
\hline
$L/a$ & 20 & 40 & 80 & 160 & 320 & 640 \\ 
\hline
$\beta$ & 4.5 & 5.0 & 5.55 & 6.2 & 6.75 & 7.45 \\
\hline
$\xi(L)/a$ & 8.78(1) & 16.76(1) & 32.26(3) & 64.6(1) & 123.4(2) & 253(1) \\ 
\hline
$\tau$ & 0.721(3) & 0.719(4) & 0.724(6) & 0.719(1) & 0.715(2) & --- \\
\hline
\end{tabular}
\caption{\it Numerical data for the correlation length $\xi(L)$ and the 
autocorrelation time $\tau$ for an $SU(3)$-symmetric ferromagnet corresponding 
to a $CP(2)$ model in D-theory.}
\label{table1}
\end{center}
\end{table}

We have analyzed the autocorrelation function for the magnetization which is
illustrated in figure 3.
\begin{figure}[tbh]
\begin{center}
\vspace{-0.4cm}
\epsfig{file=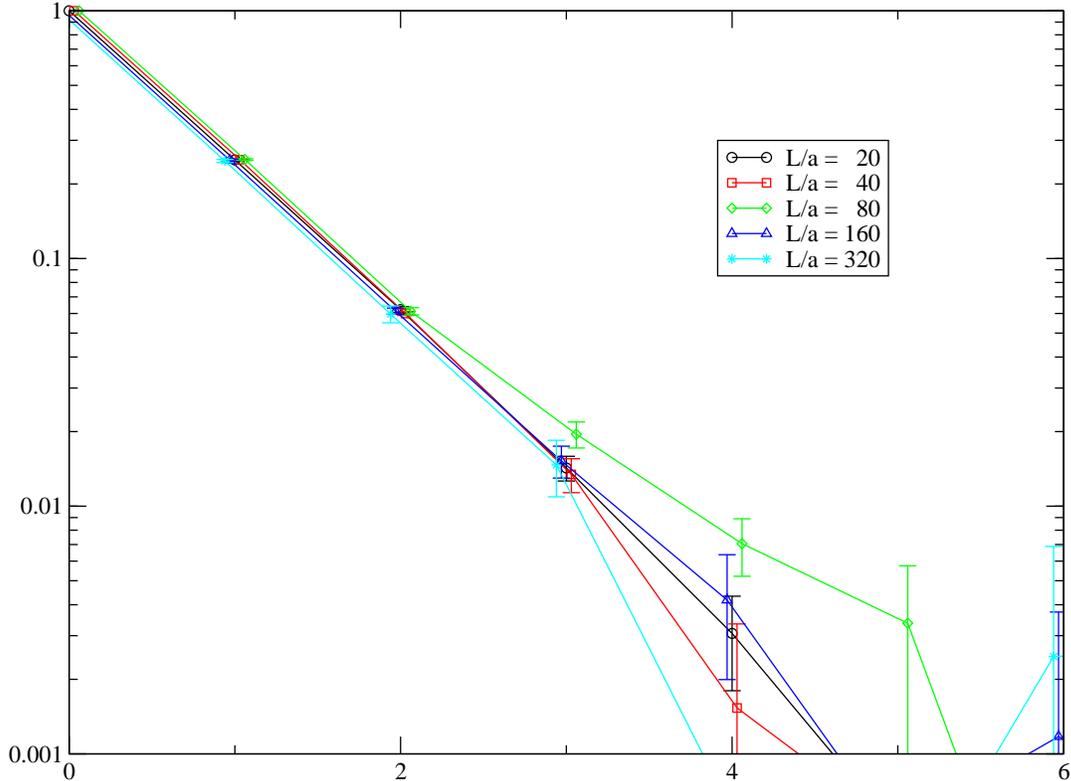,width=11.5cm,angle=-90}
\end{center}
\caption{\it Autocorrelation function of the magnetization of an $SU(3)$ 
ferromagnet as a function of the number of sweeps with the multi-cluster 
algorithm. In order to increase visibility, some points have been slightly 
displaced.}
\end{figure}
In order to estimate the dynamical exponent $z$ of critical slowing down, we 
have considered several lattices with decreasing lattice spacing, keeping the 
ratio $y=\xi(L)/L$ of the correlation length $\xi(L)$ (obtained with the
continuous time cluster algorithm using the second moment method) and the size
$L$ fixed. Our data are summarized in table 1. Since the autocorrelation time
$\tau$ does not show any dependence on $\xi(L)$, critical slowing down is
practically eliminated. In particular, our data are consistent with $z = 0$.

\section{Conclusions}

We have constructed an efficient cluster algorithm for $CP(N-1)$ models within
the theoretical framework of D-theory. The continuous classical $CP(N-1)$ 
field then emerges dynamically from the dimensional reduction of discrete 
$SU(N)$ quantum spins. Our cluster algorithm is a generalization of the loop
algorithm for $SU(2)$ quantum spins. We have investigated the algorithm's 
efficiency by estimating its autocorrelation times. Our Monte Carlo data show 
that critical slowing down is practically absent, i.e.\ the dynamical critical 
exponent is consistent with $z = 0$. This is thus another example where the 
D-theory formulation
allows us to construct efficient numerical methods which are not available 
using the standard Wilson lattice field theory. In particular, the use of the
discrete D-theory variables evades the no-go theorem that prevents the
construction of efficient Wolff-type embedding algorithms for $CP(N-1)$ models
in the standard Wilson formulation. Interestingly, using spin ladders with both
ferro- and antiferromagnetic couplings, $CP(N-1)$ models can even be simulated 
at vacuum angle $\theta = \pi$ \cite{Bea04} with no additional numerical 
effort. In particular, the sign problem that afflicts the standard approach 
is completely eliminated. Using the cluster algorithm described in this paper
one can perform accurate numerical simulations of the $CP(N-1)$ mass gap for
which no analytic results are available. Also the approach to the large $N$ 
limit can now be investigated in detail. It is also interesting to ask if 
D-theory allows us to construct efficient cluster algorithms for other field 
theories such as, e.g., $SU(N) \otimes SU(N)$ chiral models with $N \geq 3$ to 
which the no-go theorem also applies. Of course, the most challenging 
application of D-theory would be to non-Abelian gauge theories and, in 
particular, to QCD \cite{Cha97,Bro99,Bro04}.

\section*{Acknowledgments}

This work was supported in part by the Schweizerischer Nationalfonds.

\end{document}